\documentclass[a4paper,12pt]{article}
\usepackage{amsmath}
\usepackage{amsfonts}
\usepackage[vcentermath,enableskew]{youngtab}

\newcommand{\R}{{\bf{R}}}
\newcommand{\C}{{\bf{C}}}
\newcommand{\Z}{{\bf{Z}}}
\newcommand{\CP}{{{\bf CP}}}

\newcommand{\calA}{{\mathcal{A}}}
\newcommand{\calS}{{\mathcal{S}}}
\newcommand{\calP}{{\mathcal{P}}}

\newcommand{\calM}{{\mathcal{M}}}
\newcommand{\dd}{{\mathrm{d}}}
\newcommand{\ee}{{\mathrm{e}}}
\newcommand{\ii}{{\mathrm{i}}}
\newcommand{\F}{{\mathrm{F}}}
\newcommand{\T}{{\mathrm{T}}}

\newcommand{\tr}{\operatorname{tr}}
\newcommand{\spanop}{\operatorname{span}}

\newcommand{\Pchar}{{{P}}}
\newcommand{\alg}{\R[S_j]}

\newcommand{\ket}[1]{\left|#1\right\rangle}
\newcommand{\bra}[1]{\left<#1\right|}
\newcommand{\ketbra}[1]{\ket{#1}\bra{#1}}
\newcommand{\leftpar}{\overleftarrow{\partial}\!}
\newcommand{\rightpar}{\overrightarrow{\partial}\!}
\newcommand{\stab}{S[U(k)\times U(N-k)]}
\newcommand{\stabii}{S[U(2)\times U(N-2)]}
\newcommand{\eq}{eq.}
\newcommand{\eqs}{eqs.}

\newlength{\tinybox}
\Yboxdim\tinybox

\begin{document}
\begin{titlepage}
\begin{flushright}
DIAS-STP-01-16\\
hep-th/0111020\\
October 2002
\end{flushright}

\vspace{5mm}
%\vspace{1mm}

\begin{center}
  \textbf{\Large Fuzzy Complex Grassmannian Spaces\\[.5em]
    and their Star Products}

  \vspace{1cm}

  {Brian P.~Dolan
    \footnote{bdolan@thphys.may.ie}
    $\!\!{}^{,a}$
    and
    Oliver Jahn
    \footnote{jahn@stp.dias.ie}
    $\!\!{}^{,b}$}
  \\[4ex]
  {${}^a${\it 
      Department of Mathematical Physics\\
      NUI Maynooth, Maynooth, Ireland}\\[1em]
    ${}^b${\it 
      Dublin Institute for Advanced Studies\\
      10 Burlington Road, Dublin 4, Ireland}\\[1ex]}
\end{center}

\vspace{2.5mm}

\begin{abstract}
  We derive an explicit expression for an associative star product on
  non-commutative versions of complex Grassmannian spaces, in
  particular for the case of complex 2-planes.  Our expression is in
  terms of a finite sum of derivatives.  This generalises previous
  results for complex projective spaces and gives a discrete
  approximation for the Grassmannians in terms of a non-commutative
  algebra, represented by matrix multiplication in a finite-dimensional
  matrix algebra.  The matrices are restricted to have a dimension
  which is precisely determined by the harmonic expansion of functions
  on the commutative Grassmannian, truncated at a finite level.  In the
  limit of infinite-dimensional matrices we recover the commutative
  algebra of functions on the complex Grassmannians.
\end{abstract}

\end{titlepage}

\section{Introduction}

There has recently been much interest in non-commutative geometry,
\cite{Connes,Madore}, both as a novel direction in string theory
\cite{SW} and as a new tool in quantum field theory \cite{GS}.  In the
latter approach the theory is formulated on a ``fuzzy'' space, a
series of discrete approximations to a continuous space-time manifold
$\calM$.  The space of fields on each approximation is
finite-dimensional, so the fuzzy space serves as a regulator,
similar to a lattice.  In contrast to the latter, the truncation
enjoys all symmetries of the approximated continuous manifold.
%This is only possible at the expense of making the product of
%functions non-commutative.  
This has several useful implications including the absence of a
doubling problem for chiral fermions
\cite{Carow-Watamura,no-doubling}.  Topologically non-trivial field
configurations can be included and the chiral anomaly emerges
naturally in this approach \cite{topology1,topology2}.

In more detail, a fuzzy space is given by a series of
%In the latter approach the theory is formulated on a discrete
%approximation to a continuous manifold $\calM$, a ``fuzzy'' space,
%which is constructed in such a way as to preserve the symmetries of
%the manifold.  A fuzzy space is given by a series of
finite-dimensional algebras $A_L$ that approximate the commutative
algebra of functions on $\calM$ in the following way: each $A_L$ is
identified with a subset of functions on $\calM$ and the product in
$A_L$ induces a (non-commutative) product of functions in this subset,
the so-called star product; in the limit $L\to\infty$, the subset
should exhaust the set of all functions and the star product go over
to the (commutative) pointwise product of functions.  In the present
case, the algebras $A_L$ are full matrix algebras.  The fuzzy spaces
are thus matrix geometries, which go over to the usual continuous
manifold as the size of the matrix is taken to infinity.

Non-commutative star products are known to exist for every Poisson
manifold, in particular for symplectic manifolds \cite{Kontsevich}.
Star products realised by finite-dimensional matrix algebras have been
constructed in \cite{Berezin} for all homogeneous K\"ahler manifolds
provided a certain quantisation condition on the metric is satisfied.
These algebras can also be obtained using generalised coherent states
\cite{Perelomov} or the method of orbits for irreducible
representations of Lie groups \cite{Kostant,Kirillov}.  The relation
between these approaches has been discussed in \cite{RCG} and that to
deformation quantisation in \cite{CGR,Karabegov}.  For the relation to
Fourier transformation on group space see \cite{Varilly}.  In terms of
functions on the manifold, the above formulations provide an
expression of the star product as an integral over the analytical
continuation of the functions.  This is not very convenient for
explicit calculations in non-commutative field theory.  An explicit,
local formula in terms of a finite number of derivatives is so far
only known for complex projective spaces \cite{CPN} including the case
of the fuzzy 2-sphere \cite{Madore-sphere} already treated in
\cite{Presnajder}.

In this paper, we derive an analogous formula for fuzzy complex
Grassmannians, that is finite matrix geometry approximations to the
usual Grassmannians, $G^N_k\cong U(N)/[U(k)\times U(N-k)]$, which are
homogeneous spaces isomorphic to the space of complex $k$-planes in
${\bf C}^N$.  The matrix geometries consist of matrices acting on the
irreducible representation of $SU(N)$ which is given by the $L$-fold
Young product of the representation on antisymmetric $k$-tensors.  As
$L$ is increased the continuous Grassmannian is recovered.  The
special case $k=1$ requires the $L$-fold symmetric product of the
fundamental representation of $SU(N)$, as in \cite{CPN}.  The star
product for the case $k=2$ is constructed explicitly, using globally
well-defined but over-complete co-ordinates on the Grassmannian.  The
result is expressed as a finite sum over multiple derivatives of the
functions, which are decomposed into irreducible representations of
the stability group $S[U(k)\times U(N-k)]$ acting on the tangent
space.  It is shown that the star product reduces to the usual
commutative product as $L\rightarrow\infty$.  The particular case
$G^4_2$ may be of some interest in field theory as it is a symplectic
manifold which has $S^4$ as a Lagrangian sub-manifold.  The
corresponding matrix geometries can be viewed as non-commutative
versions of $T^*\!S^4$, the co-tangent bundle to $S^4$ \cite{PP}.  A
star product on the complex Grassmannians as an infinite sum over
derivatives is known \cite{Schirmer,Bordemann}.  However, this formula
cannot be restricted to finite-dimensional sub-algebras and therefore
cannot serve as a star product on a fuzzy approximation to the
manifold.

The layout of the paper is as follows: in section~\ref{sec:grassmann}
we describe the complex Grassmannians, $G^N_k$, in terms of projection
operators acting in ${\bf C}^N$, we introduce a set of global
co-ordinates which are over-complete and satisfy a set of quadratic
constraints which ensures that they indeed describe $G^N_k$; in
section~\ref{sec:harmonic} we analyse the algebra of functions in terms
of representations of $SU(N)$; in section~\ref{sec:matrix} we describe
the finite matrix geometries that define the fuzzy Grassmannians
$G^N_{k,\F}$; section~\ref{sec:star} gives the construction of the star
product for the special case $k=2$, $G^N_{2,\F}$, while
section~\ref{sec:observations} presents a conjecture on the possible
form for $k>2$ and some observations on the construction;
section~\ref{sec:conclusions} gives a summary and conclusions and some
technical details are relegated to the appendices.

\section{Complex Grassmannian spaces}
\label{sec:grassmann}

The complex Grassmannian space $G^N_k\cong U(N)/[U(k)\times
U(N-k)]=SU(N)/\stab$ can be represented as the space of all Hermitean
rank-$k$ projectors $\calP$ acting on $\C^N$.  This is easily seen
since any such projector can be diagonalised by an element of $U(N)$
and there is still a residual adjoint action of $U(k)\times U(N-k)$
which leaves it invariant.  It will be convenient to describe $G^N_k$
using a redundant set of globally well-defined co-ordinates, $\xi^A$,
$A=1,\ldots,N^2-1$.  First we introduce orthonormal Hermitean
generators $t_A$ of the Lie algebra of $SU(N)$ satisfying
\begin{equation}
  \label{eq:t-algebra}
  t_A t_B = \tfrac1N \delta_{A B} + \tfrac1{\sqrt2} ( d_{A B}{}^C + i
  f_{A B}{}^C ) t_C\;,
\end{equation}
where $f_{AB}{}^C$ are the structure constants of $SU(N)$ and
$d_{AB}{}^C$ the components of the usual symmetric traceless tensor.
The $t_A$ are normalised so that $\tr(t_At_B)=\delta_{AB}$.  This
allows us to parameterise $\calP$ in terms of $N^2-1$ real parameters
$\xi^A$,
\begin{equation}
  \label{eq:xi}
  \calP = \tfrac k N + \xi^A t_A \;.
\end{equation}
The condition that $\calP$ is as projector translates into
\begin{equation}
  \label{eq:constraints}
  \xi^A \xi^A = \frac{k(N-k)}{N}
  \qquad\text{and}\qquad
  \frac1{\sqrt2} d_{A B}{}^C \xi^A \xi^B = \frac{N-2k}{N} \, \xi^C \;,
\end{equation}
and $\xi^A$ now parameterise $G^N_k$ when these conditions are
imposed.  This construction embeds the Grassmannian in the space of
traceless Hermitian matrices, which we identify with ${\bf R}^{N^2-1}$
parameterised by the unrestricted $\xi^A$.

As discussed in detail in reference \cite{CPN} for the case of
$\CP^{N-1}\equiv G_1^N$, the complex structure and metric are encoded
in a Hermitean projector
\begin{equation}
  \label{eq:K}
  K^A{}_B \equiv \tr \bigl[ \calP t^A (1-\calP) t_B \bigr]
  = \tfrac12 \bigl( P^A{}_B + i J^A{}_B \bigr)
\end{equation}
where $J^A{}_{\!B}=\sqrt2f^{A}{}_{\!B C} \xi^C$ is the complex
structure on $G^N_k$ and $P^A{}_{\!B}=P_B{}^{A}=-(J^2)^A{}_{\!B}$
(indices are raised and lowered with the flat metric $\delta^A_B$ of
$\R^{N^2-1}$).
The matrix $K^A{}_B$ projects derivatives with respect to $\xi^A$ onto
the holomorphic tangent space of the Grassmannian when acting on the
right and onto the anti-holomorphic tangent space when acting on the
left; so $\nabla_A\equiv K_A{}^{B}\partial/\partial\xi^B$ is a
holomorphic derivative and $\bar\nabla_B\equiv
K^{A}{}_{B}\partial/\partial\xi^A$ an anti-holomorphic one.
To see that $K^A{}_{B}$ is a projector we use the completeness
relation for the generators,
\begin{equation}
  \label{eq:t-completeness}
  {(t_A)^i}_j{(t^A)^k}_l
  = {\delta^i}_l{\delta^k}_j-\tfrac{1}{N}{\delta^i}_j{\delta^k}_l,
\end{equation}
to show that 
\begin{equation}
  \label{eq:kaytee}
  K^A{}_B t^B = \calP t^A (1-\calP)
  \qquad \hbox{and}\qquad
  t_B K^B{}_A = (1-\calP) t_A\calP.
\end{equation}
(These relations will prove very useful in the ensuing analysis.)
Examining the real and imaginary parts of $K^{A}{}_{B}$ separately
reveals that $P=-J^2$ and $PJ=JP=J$.  This means that $P$ itself is a
projector onto the tangent space of $G_k^N$, with rank $2k(N-k)$.
An important observation for the following analysis is that the
differential operators $K_{A}{}^{B}\partial/\partial\xi^B$ commute with
the constraints (\ref{eq:constraints}), as they must do, since $K$
projects onto the tangent space.  This can also be proven using
(\ref{eq:kaytee}).  It is this fact that allows us to use the global
co-ordinates, rather than local co-ordinates, in the final differential
expression for the star product.

Covariant derivatives can be constructed by projecting derivatives
with respect to the flat coordinates $\xi^A$ to the tangent space.
Multiple covariant holomorphic derivatives are thus defined as
\begin{equation}
  \label{eq:multi-hol-def}
  \nabla_{A_1} \cdots \nabla_{A_n} f(\xi)
  \equiv \bigl( K_{A_1}{}^{B_1} \cdots K_{A_n}{}^{B_n} \bigr)
  \bigl( \partial_{B_1} K_{B_2}{}^{C_2}\partial_{C_2} \cdots 
  K_{B_n}{}^{C_n} \partial_{C_n} f(\xi) \bigr) 
\end{equation}
where $\partial_A=\partial/\partial\xi^A$.  In our case there is a
simplification because
\begin{equation}
  \label{eq:KKdK}
  K^{A B} K^{C D} (\partial_B K_D{}^E) = 0
\end{equation}
which follows from the definition (\ref{eq:K}) of $K$ and the
completeness relation (\ref{eq:t-completeness}).  It implies
\begin{equation}
  \label{eq:multi-hol}
  \nabla_{A_1} \cdots \nabla_{A_n} f(\xi)
  = \bigl( K_{A_1}{}^{B_1} \cdots K_{A_n}{}^{B_n} \bigr)
  \bigl( \partial_{B_1} \cdots \partial_{B_n} f(\xi) \bigr) \;.
\end{equation}

\section{Harmonic analysis}
\label{sec:harmonic}

In order to obtain a finite-dimensional truncation of the space of
functions on the Grassmannian, which is compatible with the symmetries,
we decompose the space of functions into irreducible representations
of the isometry group $G=SU(N)$.

We think of $G^N_k$ as the space $G/H$ of (right) cosets in $G$ with
respect to $H=\stab$, the subset of matrices in $U(k)\times U(N-k)$
with unit determinant.  Functions on $G/H$ can thus be considered
as functions on $G$ that are invariant under the left action of $H$.
They transform under $G$ according to the right action.

The full space of functions on $G$ is spanned by the matrix elements
$D^J_{M M'}$ of all irreducible unitary representations $J$ of $G$.
Decomposing into irreducible representations of $H$, we may write the
first component index as $M=(n,j,m)$ where $j$ labels the irreducible
representations of $H$, $m$ the corresponding components and $n$
distinguishes copies of equivalent representations.  The left action
of $H$ is then given by
\begin{equation}
  \label{eq:right-action}
  D^J_{(n,j,m)\,M'}(h^{-1}g) 
  = \sum_{m''} D^J_{(n,j,m)\,(n,j,m'')}(h^{-1}) \,
  D^J_{(n,j,m'')\,M'}(g) \;.
\end{equation}
The $H$-invariant matrix elements are those for which the first index 
corresponds to the trivial representation of $H$, labelled for
instance by $j=m=0$.  The space of functions on $G/H$ is thus spanned
by the matrix elements $D^J_{(n,0,0)\,M'}$.  

Under the right action of $G$,
\begin{equation}
  \label{eq:left-action}
  D^J_{(n,0,0)\,M'}(g g') =
  \sum_{M''} D^J_{(n,0,0)\,M''}(g) \, D^J_{M''\,M'} (g') \;,
\end{equation}
so, for fixed $J$ and $n$, the $D^J_{(n,0,0)\,M'}$ span the vector
space of the representation $J$ of $G$.  The space of functions on
$G_k^N$ thus contains all irreducible representations of $SU(N)$ that
contain the trivial representation upon restriction to $\stab$.  The
multiplicities are given by the multiplicity of the trivial
representation in the restriction.

We will describe representations of $SU(N)$ by Young diagrams.  These
will be denoted by their symbols $J=[j_1,j_2,\ldots,j_{N-1}]$ where
$j_i$ denotes the number of \emph{columns} of height $i$ of the
diagram.\footnote{Note that this symbol is different from the highest
  weight vector sometimes used to describe a diagram, the entries of
  the latter being the number of boxes in each row.}  The fundamental
representation, for instance, has $J=[1,0,\dots]$ and the adjoint one
$J=[1,0,\dots,0,1]$.  Note that the complex conjugate representation
is given by $J^*=[j_{N-1},\dots,j_1]$.  In appendix
\ref{app:representations}, we show that a representation $J$ of
$SU(N)$ contains the trivial representation of $\stab$ if and only if
it appears in the direct product
\begin{equation}
  \label{eq:matrix-alg}
  M_L \equiv [\mathbf{0}_{k-1},L,\mathbf{0}_{N-k-1}] \otimes
  [\mathbf{0}_{N-k-1},L,\mathbf{0}_{k-1}]
\end{equation}
for $L\ge n/N$ where $n=\sum_i j_i$ is the number of boxes in the
diagram $J$ and $\mathbf{0}_{k-1}$ stands for $k-1$ zero entries.  The
multiplicity in the restriction is the same as that in the direct
product.  The $M_L$ satisfy $M_1\subset M_2\subset\cdots$, so they
provide a hierarchy of truncations of the space of functions on the
Grassmannian.

The representation $[\mathbf{0}_{k-1},L,\mathbf{0}_{N-k-1}]$ is the
Young product of $L$ anti-symmetric $k$-tensors, for instance
$\yng(5,5)$ for $k=2$, $L=5$.  As an example, for $N=6$ and $k=2$, the
first truncation is
\begin{equation}\arraycolsep.166667em
  \label{eq:matrix-1}
  \begin{matrix}
    M_1 = & [0,1,0,0,0] & \otimes & [0,0,0,1,0]
    & = & [0,0,0,0,0] & \oplus & [1,0,0,0,1] & \oplus & [0,1,0,1,0]
    \rlap{\;,}
    \\ \noalign{\medskip}
    & \yng(1,1) & \otimes & \yng(1,1,1,1) \;\;
    & = & 1 & \oplus & \yng(2,1,1,1,1) & \oplus & \yng(2,2,1,1) 
  \end{matrix}
\end{equation}
the second is
\begin{equation}\arraycolsep.166667em
  \label{eq:matrix-2}
  \begin{matrix}
    M_2 = & [0,2,0,0,0] &\otimes& [0,0,0,2,0]
    &=& [0,0,0,0,0] &\oplus& [1,0,0,0,1] &\oplus& [0,1,0,1,0] \\
    \noalign{\medskip}
    &&&& \oplus& [2,0,0,0,2] &\oplus& [1,1,0,1,1] &\oplus& [0,2,0,2,0]
    \rlap{\;.}
    \\ \noalign{\medskip}
    & \yng(2,2) &\otimes& \yng(2,2,2,2)
    &=& 1 &\oplus& \yng(2,1,1,1,1) &\oplus& \yng(2,2,1,1) \\
    \noalign{\medskip}
    &&&&\oplus& \yng(4,2,2,2,2)
    &\oplus& \yng(4,3,2,2,1)
    &\oplus& \yng(4,4,2,2) 
  \end{matrix}
\end{equation}

Although not needed in the following, we would like to state for
illustrational purposes that the full harmonic analysis on $G^N_k$
(the ``union'' of all $M_L$) is given by the representations
\begin{equation}
  \label{eq:harmonic}
  J = \begin{cases}
    [m_1,\ldots,m_k,0,\ldots,0,m_k,\ldots,m_1] & \text{if } 2k<N
    \;, \\{}
    [m_1,\ldots,m_{k-1},2m_k,m_{k-1},\ldots,m_1] & \text{if } 2k=N \;,
  \end{cases}
\end{equation}
with $m_i=0,1,\dots$, each representation occurring once.  The case
$2k>N$ can be obtained by replacing $k$ by $N-k$, since $G^N_k\cong
G^N_{N-k}$.  The representations (\ref{eq:harmonic}) correspond to
Young diagrams which can be obtained by putting a diagram with at most
$k$ rows next to its conjugate (which has at least $N-k\ge k$ boxes in
any column), as can be verified for the examples given above.  This
result is also derived in appendix \ref{app:representations}.

\section{Matrix geometry}
\label{sec:matrix}

In the previous section, we have obtained a series of truncations of
the space of functions on the Grassmannian.  We will now see that these
carry a natural product.  The representation content $M_L$ of a given
truncation can in fact be realised as an algebra of matrices in a
representation $J$ of $SU(N)$: since such matrices transform under
$SU(N)$ by conjugation, they form the representation space of $J\otimes
J^*$; so $M_L$, as introduced in \eq~(\ref{eq:matrix-alg}), is
equivalent to the space of matrices in the representation
$[\mathbf{0}_{k-1},L,\mathbf{0}_{N-k-1}]$.  Since the matrix product
respects the action of $SU(N)$, the algebra $M_L$ has the same
symmetries as $G^N_k$.

In order to obtain the corresponding product of (truncated) functions,
we shall now construct an injective map from $M_L$ to the space of
functions on the Grassmannian which also respects the group action (an
equivariant map).  This map automatically provides the notions of
differentiation and integration needed for the construction of actions:
one just has to map the corresponding notions for functions back to
matrices.  Equivariance guarantees that they are compatible with the
truncation.  The map will also provide a non-commutative product for
functions in the image of $M_L$, the star product.  If the star product
tends to the point-wise product in the limit $L\to\infty$, we have
succeeded in constructing a fuzzy $G_k^N$.  Since we will restrict
ourselves to $k=2$ in the following sections, we present the map only
for this case.  The generalisation to other values of $k$ should be
obvious.

For $G^N_2$ the basic building block will be the anti-symmetric
representation $\yng(1,1)$\;, corresponding to $L=1$.  The first
non-trivial truncation of functions therefore requires using
$[N(N-1)/2]\times [N(N-1)/2]$ matrices.  A function on $G^N_2$ is
associated with such a matrix $\hat F$ by restricting the tensor
product of the fundamental projector (\ref{eq:xi}) to the anti-symmetric
representation ${\yng(1,1)}$,
\begin{equation}\label{eq:rho}
  \rho\equiv(\calP\otimes\calP)_{a} \;, 
\end{equation}
and constructing
\begin{equation}
  F_1(\xi) = \tr[\rho(\xi) \hat F] \;.
\end{equation}
Since $\calP$ has rank 2, $\rho$ has rank 1: let the plane onto which
$\calP$ projects be spanned by the vectors $\vec v$ and $\vec w$;
$\rho$ then projects onto the 1-dimensional subspace of the
representation space of ${\yng(1,1)}$ spanned by the anti-symmetric
product of $\vec v$ and $\vec w$ (an explicit proof is given in
appendix \ref{app:states}).  A general truncation requires taking the
$L$-fold (Young) product
$[0,L,0,\dots]=\yng(2,2)\genfrac{}{}{0pt}{}{\cdots}{\cdots}\yng(1,1)$
of $\yng(1,1)$\;, which has dimension
$n^N_L=\frac{(N+L-1)!(N+L-2)!}{(N-1)!L!(N-2)!(L+1)!}$, and using
$n^N_L\times n^N_L$ matrices.  Equations (\ref{eq:matrix-1}) and
(\ref{eq:matrix-2}), for instance, show the decomposition of
$15\times15$ and $105\times105$ matrices as harmonics for $G^6_2$.  In
the following we shall drop trailing zeros in symbols of
representations, so the above representations will be denoted by
$[0,L]$.  The Young product can be obtained as a component of the
symmetric tensor product, so a projector can be constructed by
restricting the $L$-fold tensor product of $\rho$ to the representation
$[0,L]$,
\begin{equation}
  \label{eq:rho-L}
  \rho_L = 
  (\,\overbrace{{\rho\otimes\cdots\otimes\rho}}^{L\ \text{times}}\,)%
  _{[0,L]} \;,
\end{equation}
(of course $\rho_1=\rho$) and a function can be associated with any
$n^N_L\times n^N_L$ matrix $\hat F$ by
\begin{equation}
  \label{eq:map}
  F_L(\xi) = \tr[\rho_L(\xi) \hat F] \;.
\end{equation}
Appendix \ref{app:states} contains a proof that the map (\ref{eq:map})
is injective.

The matrix geometries introduced here coincide with those obtained
from complex line bundles \cite{Berezin} or generalised coherent
states \cite{Perelomov}, see \cite{RCG,CGR}.  $F_L$ is usually
called the covariant symbol of the operator $\hat F$ in these
formulations, and injectivity is well known and follows from an
analyticity argument.  The relation with coherent states will be
discussed in some more detail in appendix \ref{app:states}.
%The relation to geometric quantisation has been discussed in
%\cite{RCG} and that to deformation quantisation in .  

\section{Star product on $G_2^N$}
\label{sec:star}

Multiplication of truncated functions on the Grassmannian can now be
defined using matrix multiplication.  The star product of two
functions, $F_L=\tr(\rho_L\hat F)$ and $G_L=\tr(\rho_L\hat G)$, is
obtained from the matrix product through the map
\begin{equation}
  \label{eq:star}
  (F_L\star G_L)(\xi) = \tr\bigl[ \rho_L(\xi) \hat F \hat G \bigr] \;.
\end{equation}
By construction this is an associative product and it keeps within the
class of functions truncated at level $L$.  Our aim is to find an
explicit expression for this star product, purely in terms of $F_L$ and
$G_L$ and their derivatives, thus eliminating the explicit reference
to matrices.

By orthonormality of the matrix elements in the representation
$[0,L]$,\break $\int\dd\mu(g)\, D^{[0,L]}_{M_1 M_2}(g^{-1})
D^{[0,L]}_{M_3 M_4}(g) = (1/n_L^N)\delta_{M_1 M_4}\delta_{M_2 M_3}$,
$\hat F$ can be expanded as
\begin{equation}
  \label{eq:D-expansion}
  \hat F = \int\dd\mu(g) \, \tilde F(g) D^{[0,L]}(g)
\end{equation}
with
\begin{equation}
  \label{eq:F-tilde}
  \tilde F(g) \equiv n_L^N \tr\bigl[ D^{[0,L]}(g^{-1}) \hat F \bigr] \;.
\end{equation}
Inserting this into (\ref{eq:map}), we obtain
\begin{equation}
  \label{eq:F-omega}
  F_L(\xi) = \int\dd\mu(g) \, \omega_L(\xi,g) \tilde F(g)
\end{equation}
with
\begin{equation}
  \label{eq:omega}
  \omega_L(\xi,g) \equiv \tr\bigl[ \rho_L(\xi) D^{[0,L]}(g) \bigr]
\end{equation}
and the star product can be expressed as
\begin{equation}
  \label{eq:star-omega}
  (F_L\star G_L)(\xi) = \int\dd\mu(g) \int\dd\mu(g')\,
  \omega_L(\xi,g g') \tilde F(g) \tilde G(g') \;.
\end{equation}
We seek an expression for the star product in terms of derivatives
acting on $F_L(\xi)$ and $G_L(\xi)$.  By \eqs~(\ref{eq:star-omega})
and (\ref{eq:F-omega}), this can be achieved by deriving an expression
for $\omega_L(\xi,g g')$ in terms of derivatives of $\omega_L(\xi,g)$
and $\omega_L(\xi,g')$ with respect to $\xi$.  The latter is greatly
facilitated by the observation that $\omega_L$ can be expressed
in terms of $\omega_1$,
\begin{equation}
  \label{eq:omega-L-is-power}
  \omega_L(\xi,g) = [\omega_1(\xi,g)]^L \;,
\end{equation}
because, by \eq~(\ref{eq:rho-L}), $\rho_L$ factorises into rank-1
projectors $\rho$ and $D^{[0,L]}$ acts as a direct product as well.
The reason behind this relation is that the representation $[0,L]$ when
projected to $\stabii$ by $\rho_L$ factorises as $[0,1]^L$ since all
other irreducible components of the product involve tensors that are
anti-symmetric in 3 or more indices and therefore vanish in $SU(2)$.

As a first step, we have to find an expression for $\omega_1(\xi,g
g')$.  This is a straightforward but somewhat lengthy exercise.  It
is deferred to appendix \ref{app:omega} and yields
\begin{equation}
  \label{eq:star-1}
  \omega_1(\xi,g g')=
  \omega_1(\xi,g) \left(
    1 + \leftpar_A K^{A B} \rightpar_B
    + \tfrac14 \leftpar_A \leftpar_B K^{A C} K^{B D}
    \rightpar_C \rightpar_D
    \right) \omega_1(\xi,g')
\end{equation}
where $\partial_A=\partial/\partial\xi^A$ and $K$ is the projector onto
the holomorphic tangent space introduced in \eq~(\ref{eq:K}).
Substituting (\ref{eq:star-1}) in (\ref{eq:star-omega}) and
interchanging differentiation with respect to $\xi$ with integration
over $g$ and $g'$, we now have the star product at level one,
\begin{equation}
  (F_1\star G_1)(\xi)=F_1(\xi) \left(
    1 + \leftpar_A K^{A B} \rightpar_B
    + \tfrac14 \leftpar_A \leftpar_B K^{A C} K^{B D}
    \rightpar_C \rightpar_D
    \right) G_1(\xi) \;.
\end{equation}

\bigskip

For higher $L$ we have to consider $\omega_L=(\omega_1)^L$.
Equation~(\ref{eq:star-1}) implies
\begin{equation}
  \label{eq:omega-L-uneval}
  \begin{split}
    \omega_L(\xi,g g') =
    \sum_{n+m\le L}
    \frac{L!}{n!\,m!\,(L-n-m)!} \,
    & (\omega\omega')^{L-n-m} 
    \left[ (\partial_{A}\omega) K^{A B} (\partial_{B}\omega') \right]^n
    \\
    \times & \left[ \tfrac14
      (\partial_C\partial_D\omega) K^{C E} K^{D F}
      (\partial_E\partial_F\omega') \right]^m
  \end{split}
\end{equation}
where we have used the abbreviations $\omega\equiv\omega_1(\xi,g)$ and
$\omega'\equiv\omega_1(\xi,g')$.  The right-hand side of this equation
has to be expressed in terms of multiple derivatives acting on
$\omega_L(\xi,g)$ and $\omega_L(\xi,g')$.  It contains several
different terms with a given number of derivatives.  This means that
we have to distinguish components of multiple derivatives of $\omega_L$.

To this end, we decompose multiple holomorphic derivatives
$\nabla_{A_1}\cdots\nabla_{A_n}\omega_L$ as defined in
\eq~(\ref{eq:multi-hol}) with respect to irreducible representations of
the stability group $H$ which acts on the tangent space.  It will be
sufficient to consider the subgroup $H_0=SU(2)\times SU(N-2)$ of $H$.
Representations of $H_0$ will be denoted by $(J,J')$ where $J$ is a
representation of the first factor and $J'$ one of the second.  To find
the representation content of a single holomorphic derivative
$\nabla_A=K_A{}^B\partial_B$, note that the fundamental representation
$[1]$ of $SU(N)$ decomposes as
\begin{equation}
  \label{eq:def-decomp}
  [1]\bigr|_{H_0} = ([1],[0]) \oplus ([0],[1])
\end{equation}
into the fundamental representations of $SU(2)$ and $SU(N-2)$ upon
restriction to $H_0$.  The two components can be obtained by
projection with $\calP$ and $1-\calP$.  Now use \eqs~(\ref{eq:K}) and
(\ref{eq:t-completeness}) to write $\nabla_A$ in terms of
(anti-)fundamental indices,
\begin{equation}
  \label{eq:hol-fund}
  (t^A)^i{}_j \nabla_A f(\xi)
  = \bigl[ (1-\calP) t^B \calP \bigr]^i{}_j \, \partial_B f(\xi) \;.
\end{equation}
The matrix $t^B\partial_B f$ transforms like the traceless component
of $[1]\times[1]^*$ under $SU(N)$.  Since the index $i$ is projected
by $1-\calP$ to $([0],[1])$ while $j$ is projected by $\calP$ to
$([1],[0])^*$, we find that the holomorphic derivative transforms like
$([0],[1])\otimes([1],[0])^*=([1]^*,[1])$.  Note that tracelessness is
guaranteed by the projections in \eq~(\ref{eq:hol-fund}).  By the same
reasoning, an anti-holomorphic derivative
$\bar\nabla_A=K^B{}_A\partial_B$ transforms like $([1],[1]^*)$.  A
multiple holomorphic derivative transforms like the symmetric tensor
product of $n$ copies of the representation $([1]^*,[1])$.  In order
to obtain an explicit expression for the decomposition of this
product, it turns out to be useful to first decompose the tensor
product of $n$ copies of the fundamental representation of $SU(N)$.

This can be done by considering the action of the symmetric group
$S_n$, whose elements permute the factors in the tensor product
\cite{Sternberg}.  The latter can then be decomposed into irreducible
representations of $SU(N)\times S_n$ with the help of character
projection operators.  They provide the following decomposition of
unity,
\begin{equation}
  \label{eq:decomp-def}
  1 = \sum_{|J|=n} \Pchar_J
  \qquad\text{where}\qquad
  \Pchar_J \equiv \frac{d_J}{n!} \sum_{\pi\in S_n}
  \chi_J(\pi) \pi \;.
\end{equation}
Here, the sum is over all Young diagrams with $n$ boxes, $\chi_J$
is the character of the symmetric group in the representation $J$
and $d_J$ the dimension of that representation.  We will only need
\begin{equation}
  \label{eq:dim-sym}
  d_{[l,m]}=\frac{(l+2m)!(l+1)}{(l+m+1)!m!} \;.  
\end{equation}
The projectors $\Pchar_J$ are orthogonal, $\Pchar_J \Pchar_{J'} =
\delta_{J,J'} \Pchar_J$.  They are discussed in more detail in appendix
\ref{app:sym}.  When acting on tensors, $\Pchar_J$ projects onto a
direct product of the representations of $SU(N)$ and the symmetric
group given by $J$.  Consequently the image of $\Pchar_J$ contains
$d_J$ copies of the $SU(N)$ representation $J$.

This decomposition can be used to write the projector onto symmetric
tensors with \emph{adjoint} indices as
\begin{equation}
  \label{eq:sym-adj}
  \frac{1}{n!} \, \delta_{A_1}^{\{B_1} \cdots \delta_{A_n}^{B_n\}}
  = \frac{1}{n!} \sum_{|J|=n} \tr \bigl[ 
  \bigl( t_{A_1} \otimes \cdots \otimes t_{A_n} \bigr) 
  \bigl( t^{\{B_1} \otimes \cdots \otimes t^{B_n\}} \bigr)
  \, \Pchar_{J} \bigr] 
\end{equation}
where curly brackets denote symmetrisation.  The projector from general
adjoint tensors to symmetric holomorphic tensors (or anti-holomorphic
tensors when acting to the left) can now be decomposed as
\begin{equation}
  \label{eq:K-reduction}
  \frac{1}{n!} K_{A_1}{}^{\{B_1} \cdots K_{A_n}{}^{B_n\}}
  = \sum_{l+2m=n} K_{[l,m]}{\,}_{A_1\dots A_n}{}^{B_1\dots B_n}
\end{equation}
with projection operators
\begin{equation}
  \label{eq:K-J-explicit}
  \begin{split}
  K_J^{A_1\dots A_n,B_1\dots B_n} 
  & = \tfrac{1}{n!} \tr \bigl[ 
  \bigl( t^{A_1} \otimes \cdots \otimes t^{A_n} \bigr)
  \bigl( t^{\{C_1} \otimes \cdots \otimes t^{C_n\}} \bigr)
  \Pchar_{J} \bigr] \,
  K_{C_1}{}^{B_1} \cdots K_{C_n}{}^{B_n} 
  \\
  & =
  \tfrac{1}{n!} \tr \bigl[ 
  \bigl( t^{A_1} \otimes \cdots \otimes t^{A_n} \bigr)
  \,\Pchar_J\,
  (1-\calP)^{\otimes n}\,
  \bigl( t^{\{B_1} \otimes \cdots \otimes t^{B_n\}} \bigr)
\,  \calP^{\otimes n} 
  \,\Pchar_J \bigr]
  .
\end{split}
\end{equation}
The second expression has been obtained by inserting
\eq~(\ref{eq:kaytee}) for the contraction of $K_C{}^B$ with $t^C$ and
generating a second projector using the fact that $\Pchar_J=\Pchar_J^2$
commutes with symmetric tensors.  To see that $K_J$ is a projector, one
has to use the completeness relation (\ref{eq:t-completeness}) for the
generators $t^A$ and the fact that $\Pchar_J$ and $K_A{}^B$ are
projectors.  The first expression shows that, when acting to the right,
$K_J$ projects onto the holomorphic tangent space.  The second
expression reveals that the holomorphic tensors thus obtained are
projected to a sum of equivalent irreducible representations of
$H_0=SU(2)\times SU(N-2)$: the first indices of the $n$ matrices
$t^{B_i}$ are projected to $[1]^{\otimes n}$ of $SU(N-2)$ and then
symmetrised according to the Young diagram $J$, the second indices are
projected to ${[1]^*}^{\otimes n}$ of $SU(2)$ and symmetrised in the
same way; the result is a sum of representations equivalent to
$(J^*,J)$ of $H_0$.  Since $\Pchar_J$ projects onto $d_J$ copies of $J$
corresponding to different symmetrisations and the symmetrisation on
both sides of the symmetric $t^{\{B_1}\otimes\cdots\otimes t^{B_n\}}$
has to be the same, $K_J$ projects onto $d_J$ copies of $(J^*,J)$.
When acting to the left, $K_J$ projects onto the anti-holomorphic
tangent space since the factors $K$ commute with the trace.  The result
is projected to $d_J$ copies of the representation $(J,J^*)$.

These findings imply that $K_{[j_1,j_2,\dots]}$ vanishes if $j_i\ne0$
for any $i>2$ as can be seen explicitly by noting that the second
$\Pchar_J$ is multiplied by $\calP^{\otimes j}$ in
\eq~(\ref{eq:K-J-explicit}) and $\calP$ is a rank-2 projector.  This
fact has been used to restrict the sum in \eq~(\ref{eq:K-reduction}).
Note that $K_{[1]}^{A,B}=K^{A B}$ and
\begin{equation}
  \label{eq:K-01}
  \begin{split}
    K_{[0,1]}^{A B,C D}=
    \tfrac{1}{4}\bigl\{ &\tr\bigl[\calP t^A(1-\calP)t^C\bigr]
    \tr\bigl[\calP t^B(1-\calP)t^D\bigr]\\
    &-\tr\bigl[\calP t^A(1-\calP)t^C\calP
    t^B(1-\calP)t^D\bigr] \bigr\}
    \; + \; (C \leftrightarrow D) \;.
  \end{split}
\end{equation}

A useful observation is that $K_{[l,m]}$ factorises when contracted
with arbitrary symmetric tensors $S$ and $T$,
\begin{multline}
  \label{eq:K-factor}
  S_{\{A_1\dots A_n\}} \,
  K_{[l,m]}^{A_1\dots A_n,B_1\dots B_n} 
  \, T_{\{B_1\dots B_n\}} \\
  = d_{[l,m]} \, S_{\{A_1\dots A_n\}} \,
  K_{[l]}^{A_1\dots A_l,B_1\dots B_l}
  K_{[0,1]}^{A_{l+1} A_{l+2},B_{l+1} B_{l+2}} \cdots
  K_{[0,1]}^{A_{n-1} A_{n},B_{n-1} B_{n}} 
  \, T_{\{B_1\dots B_n\}} \;,
\end{multline}
or, in index-free notation, $S K_{[l,m]} T=d_{[l,m]} S (K_{[l]}\otimes
K_{[0,1]}^{\otimes m}) T$ where $d_{[l,m]}$ is given in
\eq~(\ref{eq:dim-sym}).  This equation is proved in appendix\ 
\ref{app:sym}.

\bigskip

The irreducible components of the derivatives of $\omega_L=\omega^L$
are calculated in appendix \ref{app:multi-der}; in index-free notation,
the result is
\begin{multline}
  \label{eq:multi-der}
  (\partial^{l+2m} \omega^L ) \, 
    K_{[l,m]} \,
    (\partial^{l+2m} \omega^{\prime L} ) \\
  = c_{l,m}^{(L)} \, 
  (\omega\omega')^{ L-l-m} 
  \bigl[ (\partial\omega)^l K_{[l]} (\partial\omega')^{l} \bigr]
  \left[ \tfrac14 (\partial\partial\omega) K_{[0,1]}
    (\partial\partial\omega') \right]^{m}
\end{multline}
where
\begin{equation}
  \label{eq:c-mult}
  c_{l,m}^{(L)} =
  \left( \frac{L! (L+1)!}{(L-l-m)! (L+1-m)!} \right)^2    d_{[l,m]} \;.
\end{equation}
The crucial ingredient in the derivation is the following relation
between single and double derivatives of $\omega$,
\begin{equation}
  \label{eq:d2omega=domega2}
  K^{A C} K^{B D} \omega \partial_C\partial_D \omega 
  = K_{[0,1]}^{A B,C D} \omega \partial_C\partial_D\omega 
  = 2 K_{[0,1]}^{A B,C D} \partial_C\omega \partial_D\omega \;,
\end{equation}
which can be obtained by use of \eqs~(\ref{eq:t-completeness}) and
(\ref{eq:PPP-rel}) in a somewhat tedious calculation.  In addition,
since $\omega$ is only quadratic in $\xi^A$, one has
$\partial_A\partial_B\partial_C\omega=0$.

With the tools provided, it is easy to bring the terms in the
expression (\ref{eq:omega-L-uneval}) for $\omega_L(\xi,g g)$ into the
form (\ref{eq:multi-der}).  Owing to \eq~(\ref{eq:d2omega=domega2}), we
have
\begin{equation}
  \tfrac14 \partial_C\partial_D\omega K^{C E} K^{D F}
  \partial_E\partial_F\omega' 
  = \partial_C\partial_D\omega K_{[0,1]}^{C E,D F}
  \partial_E\partial_F\omega' \;.
\end{equation}
Now we use the decomposition (\ref{eq:K-reduction}) together with the
factorisation property (\ref{eq:K-factor}) to write
\begin{equation}
  \label{eq:domega-reduction}
  (\partial_A\omega K^{A B} \partial_B\omega')^n =
  \sum_{l+2i=n}  d_{[l,i]} \, 
  \bigl[ (\partial\omega)^l K_{[l]} (\partial\omega')^l \bigr]
  \left[
    (\partial\partial\omega) K_{[0,1]}
    (\partial\partial\omega') \right]^i
  (\omega \omega')^i 
  \;.
\end{equation}
Inserting this into \eq~(\ref{eq:omega-L-uneval}) and combining
equivalent terms, we obtain
\begin{equation}
  \label{eq:omega-L-sum2}
  \omega_L(\xi,g g') = \sum_{l+m\le L}
  \tilde c^{(L)}_{l,m} \,
  (\omega\omega')^{L-l-m}
  \bigl[ (\partial\omega)^l  K_{[l]} (\partial\omega')^l) \bigr]
  \left[
    (\partial\partial\omega) K_{[0,1]}
    (\partial\partial\omega') \right]^m
\end{equation}
with
\begin{equation}
  \label{eq:c-sum}
  \tilde c^{(L)}_{l,m} 
  = \sum_{i=0}^{\operatorname{min}(m,L-l-m)}  
  \frac{L!}{(l+2i)! \, (m-i)! \, (L-l-m-i)!} \, d_{[l,i]} \;.
\end{equation}
The sum can be performed using the identity
\begin{equation}
  \sum_{i=0}^{\min(p,n)}
  \begin{pmatrix} p\\ i\end{pmatrix}
  \begin{pmatrix} q\\ n-i\end{pmatrix}
  = \begin{pmatrix} p+q\\ n\end{pmatrix} \;,
\end{equation}
one finds
\begin{equation}
  \label{eq:c}
    \tilde c^{(L)}_{l,m} 
    = \frac{L! (L+1)!} {(L-l-m)! (L+1-m)! (l+2m)!} \, d_{[l,m]} \;.
\end{equation}
Now the terms have exactly the form (\ref{eq:multi-der}) and we can
write
\begin{equation}
  \label{eq:omega-L-sum3}
  \omega_L(\xi,g g') = \sum_{l+m\le L} a^{(L)}_{l,m} \, 
  \bigl( \partial^{l+2m} \omega_{L}(\xi,g) \bigr)
  \, K_{[l,m]}(\xi) \,
  \bigl( \partial^{l+2m} \omega_L(\xi,g') \bigr)
\end{equation}
with
\begin{equation}
  \label{eq:a}
  a^{(L)}_{l,m} = \frac{(L-l-m)! (L+1-m)!} 
                   {L! (L+1)! (l+2m)!}\;.
\end{equation}
Inserting \eq~(\ref{eq:omega-L-sum3}) into \eq~(\ref{eq:star-omega})
and interchanging differentiation with respect to $\xi$ with
integration over $g$ and $g'$, we obtain the final expression for the
star product,
\begin{multline}
  \label{eq:star-irred}
  (F_L\star G_L)(\xi) =
  \\
  \sum_{l+m\le L}  a^{(L)}_{l,m} \, 
  \bigl( \partial_{A_1}\cdots\partial_{A_{l+2m}} F_L(\xi) \bigr)
  \, K_{[l,m]}^{A_1\dots A_{l+2m},B_1\dots B_{l+2m}}(\xi) \,
  \bigl( \partial_{B_1}\cdots\partial_{B_{l+2m}} G_L(\xi) \bigr) \;.
\end{multline}
Recall that $K_{[l,m]}$ projects the derivatives on its right to
holomorphic ones in the irreducible representation $[l,m]^*\otimes
[l,m]$ of the subgroup $SU(2)\times SU(N-2)$ of the stability group
and those on its left to anti-holomorphic ones in the complex conjugate
representation.

Since the star product (\ref{eq:star-irred}) is ultimately derived from
a matrix product it is guaranteed to be associative, by construction.
For large $L$, $a^{(L)}_{l,m}\sim L^{-l-2m}$, so all terms except
$l=m=0$ are suppressed and the star product tends to the point-wise
product.  We conclude that the series of matrix geometries introduced
in section \ref{sec:matrix} indeed constitutes a fuzzy version of the
complex Grassmannians $G_2^N$.  As a consistency check note that
restricting the sum to $m=0$ reduces this to the known result for
$\CP^{N-1}$, \cite{CPN}.

\section{Observations on the construction and generalisation to $k>2$}
\label{sec:observations}

The above results suggest a natural generalisation to the case $k>2$,
although we have not verified it: the definition of matrix geometries
in section \ref{sec:matrix} generalises immediately to Young products
of antisymmetric $k$-tensors; in \eq~(\ref{eq:star-1}), we expect
analogous terms with up to $k$ derivatives acting on each side, and
\eq~(\ref{eq:star-irred}) is expected to contain projections
$K_{[j_1\dots j_k]}$ with $\sum_i j_i\le L$; a natural extension of
the combinatorial factor (\ref{eq:a}) would be
\begin{equation}
  \label{eq:comb-k}
  a^{(L)}_J = \frac{1}{|J|!} \prod_i \frac{1}{L+r_i-c_i}
\end{equation}
where the product runs over all boxes of the Young diagram $J$ and
$r_i$ and $c_i$ are the row and column of box $i$.

\bigskip

Returning to the particular case $k=2$, it might be more convenient,
for practical purposes, to express $K_{[l,m]}$ in terms of $K$ and
$K_{[0,1]}$ alone by using the inverse of the decomposition
(\ref{eq:K-reduction}): we can add factors of $K_{[0,1]}$ on both
sides of (\ref{eq:K-reduction}) and use (\ref{eq:K-factor}) to write
\begin{equation}
  \label{eq:K-reduction-matrix}
  S \bigl( K^{\otimes(p-2n)} \otimes K_{[0,1]}^{\otimes n} \bigr) T
  = \sum_{m=n}^{[p/2]} M^{(p)}_{n,m} 
  \, S \bigl( K_{[p-2m]} \otimes K_{[0,1]}^{\otimes m} \bigr) T
\end{equation}
where $S$ and $T$ are symmetric tensors and
$M^{(p)}_{n,m}=d_{[p-2m,m-n]}$.  For fixed $p$, $M^{(p)}$ can be
considered as a triangular matrix with $M^{(p)}_{n,m}=0$ if $n>m$.  It
can be inverted by use of the relation
\begin{equation}
  \label{eq:inverse}
  \sum_{m=n}^{n'} d_{[p-2m,m-n]} \,e_{p-2n',n'-m} = \delta_{n,n'}
  \qquad\text{with}\qquad
  e_{i,j} = (-1)^j \binom{i+j}{j}
\end{equation}
which can be derived by standard techniques.  We find
$(M^{(p)})^{-1}_{m,n}=e_{p-2n,n-m}$ for $m\le n$ and 0 otherwise.  The
inverse of \eq~(\ref{eq:K-reduction-matrix}) therefore is
\begin{equation}
  \label{eq:inv-K-reduction}
  S \, K_{[p-2m,m]} \, T
  = d_{[p-2m,m]} \sum_{n=m}^{[p/2]} e_{p-2n,n-m} 
  \, S \bigl( K^{\otimes(p-2n)} \otimes K_{[0,1]}^{\otimes n} \bigr) T
\end{equation}
where the factorisation property (\ref{eq:K-factor}) has been used on
the left-hand side.  Inserting this into \eq~(\ref{eq:star-irred})
and re-expanding indices, we obtain an alternative formula for the
star product,
\begin{equation}
  \label{eq:star-red}
  \begin{split}
    (&F_L\star G_L)(\xi) = \sum_{n+m\le L}  b^{(L)}_{n,m} \;
    (\partial_{A_1}\cdots \partial_{A_{n+2m}} F_L(\xi)) \,
    \bigl( K^{A_1B_1}(\xi) \cdots K^{A_nB_n}(\xi) \\
    &\times K_{[0,1]}^{A_{n+1}A_{n+2},B_{n+1}B_{n+2}}(\xi)\cdots
    K_{[0,1]}^{A_{n+2m-1}A_{n+2m},B_{n+2m-1}B_{n+2m}}(\xi) \bigr) \,
    (\partial_{B_1}\cdots \partial_{B_{n+2m}}G_L(\xi))
  \end{split}
\end{equation}
where (\ref{eq:dim-sym}), (\ref{eq:a}) and (\ref{eq:inverse}) have
been combined into
\begin{equation}
  \label{eq:b}
  b^{(L)}_{n,m}
  =  \sum_{i=0}^{\operatorname{min}(m,L-n-m)} 
  (-1)^i
  \frac{(L-n-m-i)!\,(L+1+i-m)!\,(n+i)!\,(n+2i+1)}
       {L!\,(L+1)!\,(n+m+i+1)!\,(m-i)!\,i!\,n!} \;.
\end{equation}
$K$ and $K_{[0,1]}$ are explicitly given by
\begin{align}
  \label{eq:P-01}
  K^{A B} &= \tfrac12(-J^{A C} J_C{}^B + i J^{A B})
  \qquad\text{with}\qquad
  J^{A B}=\sqrt2 f^{A B}{}_C \xi^C \;,
  \\
  \nonumber
  K_{[0,1]}^{A B,C D} &= \left( \frac{N-2}{4N} \left(
    \delta^A_E \delta^B_F + \delta^A_F \delta^B_E \right)
  - \frac14 \left(
    d^{A G}{}_E \, d^{B}{}_{G F} + d^{A G}{}_F \, d^{B}{}_{G E} \right)
  \right)
  K^{E C} K^{F D} \;.
\end{align}

\bigskip

\section{Conclusions}
\label{sec:conclusions}

We have extended an explicit, local description of
fuzzy matrix geometries beyond the known examples of
$\CP_{\textrm{F}}^{N-1}$ to complex Grassmannians $G^N_k\cong
U(N)/[U(k)\times U(N-k)]$.  The geometry of the Grassmannian is
conveniently described in terms of $N^2-1$ globally defined
co-ordinates $\xi^A$, satisfying the quadratic constraints
(\ref{eq:constraints}), and a rank-$k$ projector $\calP$ defined in
(\ref{eq:xi}).  The fuzzy Grassmannians $G^N_{k,\F}$ are defined by
algebras of $n^N_L\times n^N_L$ matrices, $\hat F$, where $n^N_L$ is
the dimension of the irreducible representation of $SU(N)$ which is the
$L$-fold Young product of the $k$-fold anti-symmetric representation
and has Young diagram
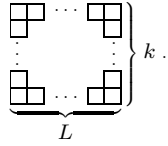
\begin{equation*}
  \hbox{\tiny\arraycolsep.25em$
  \underbrace{\begin{matrix}
    \yng(2,1) & \raisebox{.5em}{$\cdots\mkern-1mu$} & \young(\ \ ,:\ )
    \\[-.5em]
    \vdots\quad & & \quad\vdots \\
    \yng(1,2) & \raisebox{-.5em}{$\cdots\mkern-1mu$} & \young(:\ ,\ \ )
  \end{matrix}}_{\mbox{\scriptsize$L$}} 
  \! \left.\rule{0pt}{3em}\right\} \mbox{\scriptsize$k$} \;.$}
\end{equation*}
Such matrices are mapped injectively to functions using a rank-1
projector constructed from the $k$-fold anti-symmetric tensor product
of $\calP$, $\rho=(\calP\otimes\cdots\otimes\calP)_a$ as the $L$-fold
product $\rho_L=
(\rho\otimes\cdots\otimes\rho)_{[\mathbf{0}_{k-1},L,\mathbf{0}_{N-k-1}]}$.
The map is given by
\begin{equation*}
  F_L(\xi) = \tr\bigl[ \rho_L(\xi) \hat F \bigr] \;.
\end{equation*}
The set of functions thus obtained constitute a truncation of the
harmonic expansion of a general function on $G^N_k$.

An associative star product  between two such functions
can then be defined using matrix multiplication as
\begin{equation*}
  (F_L\star G_L)(\xi)=\tr\bigl[ \rho_L(\xi) \hat F \hat G\bigr] \;.
\end{equation*}
The right-hand side can be re-expressed as a (bi-)differential
operator on $F_L(\xi)$ and $G_L(\xi)$ and is given explicitly, for
$k=2$, in \eq~(\ref{eq:star-irred}),
\begin{multline*}
  (F_L\star G_L)(\xi) =
  \\
  \sum_{l+m\le L}  a^{(L)}_{l,m} \, 
  \bigl( \partial_{A_1}\cdots\partial_{A_{l+2m}} F_L(\xi) \bigr)
  \, K_{[l,m]}^{A_1\dots A_{l+2m},B_1\dots B_{l+2m}}(\xi) \,
  \bigl( \partial_{B_1}\cdots\partial_{B_{l+2m}} G_L(\xi) \bigr)
\end{multline*}
where $a^{(L)}_{l,m}$ is given by \eq~(\ref{eq:a}) and $K_{[l,m]}$ by
\eq~(\ref{eq:K-J-explicit}).  $K_{[l,m]}$ projects the multiple
derivatives on $F_L$ and $G_L$ to irreducible representations of the
stability group $H=\stabii$.  It includes a projection of the
derivatives to the tangent space of $G_2^N$ in the embedding space
parameterised by $\xi$.  Tangent vectors transform under $H$ and the
multiple tangential derivative forms a direct product representation
which is reducible.  The derivatives on $G_L$ are projected by
$K_{[l,m]}$ to the component equivalent to the irreducible
representation $[l,m]^*\otimes [l,m]$ of $SU(2)\times SU(N-2)\subset
H$.  Here, $[l,m]$ stands for a Young product of $m$ copies of the
anti-symmetric representation $\yng(1,1)$ and $l$ copies of the
fundamental representation $\yng(1)$, for instance
$[4,3]=\vcenter{\hbox{\tiny$\yng(7,3)$}}$.  The derivatives on $F_L$
are projected to the complex conjugate representation.  The projection
implies that all derivatives on $G_L$ are holomorphic while those on
$F_L$ are anti-holomorphic.  They can in fact be replaced by covariant
derivatives in local co-ordinates \cite{CPN}.  When the sum is
restricted to $m=0$ this star product reduces to the $k=1$ case of
$\CP^{N-1}$, \cite{CPN}.

An even more explicit expression for the star product, possibly better
suited for practical purposes, is given in \eq~(\ref{eq:star-red}).

Explicit, local formulas for the star product prove very useful in
perturbative calculations in non-commutative field theory \cite{phi-4}
and might also provide new insights in string theory where fuzzy
versions of co-adjoint orbits appear as world volume geometries of
D-branes in WZW models \cite{ARS,FFFS}.

\section{Acknowledgements}

The authors wish to thank A.\,P.~Balachandran, Denjoe O'Connor and
P.~Pre\v{s}najder for useful discussions and the Department of Physics,
CINVESTAV, Mexico for their hospitality during a fruitful visit.  The
work of OJ has been supported by the Alexander von Humboldt Foundation
through a Feodor Lynen Research Fellowship (grant number
V-3-FLF-1068701).

\appendix

\section{Kernel at level 1}
\label{app:omega}

We derive an expression for the kernel $\omega_1(\xi,g g')$ in terms
of derivatives of $\omega_1(\xi,g)$ and $\omega_1(\xi,g')$ with
respect to $\xi$.  Inserting \eq~(\ref{eq:rho}) and the definition of
the anti-symmetric representation, $D^{[0,1]}(g)=(g\otimes g)_a$, into
\eq~(\ref{eq:omega}) and abbreviating $\omega_1$ as $\omega$, we
obtain
\begin{equation}
  \label{eq:omega-g}
  \omega(\xi,g)
  =\tr\bigl[ (\calP\otimes\calP)_a (g\otimes g)_a \bigr]
  \qquad\text{with}\qquad
  \calP = \tfrac2N + \xi^A t_A
\end{equation}
The trace of a tensor product in the anti-symmetric representation can
be expressed as
\begin{equation}
  \label{eq:trace-01}
  \tr\bigl[ (A\otimes B)_a \bigr]
  = \tfrac12 \bigl[ \tr(A) \tr(B) - \tr(A B) \bigr] \;.
\end{equation}
A crucial identity, $\calP^i{}_{[l}
\calP^j{}_m \calP^k{}_{n]}=0$, follows from the fact that $\calP$ is a
rank-2 projector.  It implies
\begin{equation}
  \label{eq:PPP-rel}
  \begin{split}
    \tr (\calP A) \tr (\calP B) \tr (\calP C) 
    - \tr (\calP A \calP B) \tr (\calP C) 
    - \tr (\calP A \calP C) \tr (\calP B) & {}\\
    - \tr (\calP B \calP C) \tr (\calP A) 
    + \tr (\calP A \calP B \calP C)
    + \tr (\calP B \calP A \calP C) 
    &= 0 \;.
  \end{split}
\end{equation}

Since 
\begin{equation}
  \partial_A\omega = \tr\bigl[ (t_A\otimes\calP)_a(g\otimes g)_a \bigr] +
  \tr\bigl[ (\calP\otimes t_A)_a(g\otimes g)_a \bigr]
\end{equation}
we can use (\ref{eq:kaytee}) to show
\begin{equation}
  \label{eq:domega}
  \partial_A\omega(\xi,g) \, K^{A B} \, \partial_B\omega(\xi,g')
  = 2 \tr\bigl[ (\calP\otimes\calP)_a (g\otimes g)_a 
  \bigl( \calP\otimes (1-\calP) \bigr)_a
  (g'\otimes g')_a  \bigr] \;.
\end{equation}
Similarly, for the second derivative
\begin{multline}
  \label{eq:ddomega}
  \partial_A\partial_B\omega(\xi,g) \,
  K^{A C} K^{B D} \,
  \partial_C\partial_D\omega(\xi,g')
  \\
  = 4 \tr\bigl[ (\calP\otimes\calP)_a (g\otimes g)_a 
  \bigl( (1-\calP)\otimes (1-\calP) \bigr)_a
  (g'\otimes g')_a \bigr] \;.
\end{multline}
Since $\rho_1=(\calP\otimes\calP)_a$ is a rank-1 projector, the simple
product can be written as
\begin{equation}
  \label{eq:omega-prod}
  \omega(\xi,g)\omega(\xi,g')
  = \tr\bigl[ (\calP\otimes\calP)_a (g\otimes g)_a
              (\calP\otimes\calP)_a (g'\otimes g')_a \bigr] \;.
\end{equation}
Using
$1=(\calP\otimes\calP)_a+(\calP\otimes(1-\calP))_a
+((1-\calP)\otimes\calP)_a+((1-\calP)\otimes(1-\calP))_a$,
\eqs~(\ref{eq:domega}), (\ref{eq:ddomega}) and (\ref{eq:omega-prod})
can be combined to
\begin{equation}
  \label{eq:star-1-app}
  \omega(\xi,g g')=
  \omega(\xi,g) \left(
    1 + \leftpar_A K^{A B} \rightpar_B
    + \tfrac14 \leftpar_A \leftpar_B K^{A C} K^{B D}
    \rightpar_C \rightpar_D
    \right) \omega(\xi,g') \;,
\end{equation}
which is the desired expression.

\section{Coherent states}
\label{app:states}

We show that the projector $\rho_L$ as given in \eq~(\ref{eq:rho-L})
has rank 1, and we present a simple argument for why the map from
matrices to functions is injective.

To this end, we require a more explicit representation for $\rho_L$.
The vector space of the irreducible representation of $SU(N)$ with
symbol $J=[0,L]$ can be realised as a sub-space with certain
symmetry properties of the space of $2L$-index tensors.  We construct
it as the image of a Young symmetriser.  We first
assign tensor indices to the boxes in the Young diagram of the
representation by putting the numbers $1,2,\ldots,2L$ in ascending
order into one column after the other, for instance
$\vcenter{\hbox{\tiny$\young(1357,2468)$}}$ for $[0,4]$.  The Young
symmetriser is now defined as
\begin{gather}
  \label{eq:Y-0,L}
  \begin{gathered}
  Y_{[0,L]} = \frac{2^L}{L+1} \calA_L \calS_L
  \;, \\
  \calA_L = \prod_{i=1}^L \left[ \tfrac{1}{2} (1-\tau_i) \right]
  \;, \qquad
  \calS_L = \frac{1}{L!} \sum_{\pi_1\in R_1} \pi_1
  \; \frac{1}{L!} \sum_{\pi_2\in R_2} \pi_2
  \end{gathered}
\end{gather}
where $\tau_i$ interchanges the two boxes of the $i^{\textrm{th}}$
column of the diagram and $R_i$ denotes the set of permutations that
permute the boxes of row $i$.  So $\calS_L$ symmetrises the rows of the
diagrams, while $\calA_L$ anti-symmetrises the columns.  Both are
symmetric projectors.  The Young symmetriser is a projector,
$Y_{[0,L]}^2=Y_{[0,L]}$, but not symmetric.

Operators in the vector space of $[0,L]$ can be unambiguously
described as operators $\hat F$ on $2L$-tensors that satisfy
\begin{equation}
  \label{eq:F-sym}
  \hat F Y_{[0,L]} = Y_{[0,L]} \hat F = \hat F \;.
\end{equation}

Now we can prove that $\rho_L$ has rank 1 by expressing the rank-2
projector $\calP$ in terms of an orthonormal basis $\ket\varphi$,
$\ket\psi$ of the complex plane onto which it projects,
\begin{equation*}
  \label{eq:P-basis}
  \calP = \ket\varphi\bra\varphi + \ket\psi\bra\psi \;.
\end{equation*}
The level-1 projector $\rho$ was defined in (\ref{eq:rho}) as the
projection of the tensor product of $\calP$ with itself to the
anti-symmetric representation $[0,1]$.  Since $Y_{[0,1]}$ reduces to a
single anti-symmetrisation,
\begin{equation}
  \label{eq:rho-state}
  \rho \equiv (\calP\otimes\calP)_a 
  = (\calP\otimes\calP) Y_{[0,1]} = \ketbra{\varphi\psi}
\end{equation}
where
\begin{equation}
  \label{eq:state}
  \ket{\varphi\psi} \equiv \tfrac{1}{\sqrt2}
  \bigl( \ket\varphi\ket\psi - \ket\psi\ket\varphi \bigr) \;.
\end{equation}
For the projector at level $L$, we obtain
\begin{equation}
  \label{eq:PL-state}
  \rho_L = (\rho \otimes \cdots \otimes \rho) Y_{[0,L]}
  = \ket{\varphi\psi}^L \bra{\varphi\psi}^L Y_{[0,L]}
\end{equation}
The state $\ket{\varphi\psi}$ completely characterises the plane that
corresponds to a point in $G_2^N$.  It is therefore natural that it
occurs as a fundamental object in the construction.  The states
$\ket{\varphi\psi}^L$ coincide, up to a conventional phase, with the
generalised coherent states discussed in \cite{Perelomov}.  Since
$\hat F Y_{[0,L]}=Y_{[0,L]}\hat F$,
\begin{equation}
  \label{eq:map-state}
  F_L(\xi) = \bra{\varphi\psi}^L \hat F \ket{\varphi\psi}^L \;.
\end{equation}
So $F_L$ is the covariant symbol, as defined in
\cite{Perelomov}, of the operator $\hat F$.

This expression can be used to show that the map is injective.  We
have to show that $\hat F$ can be reconstructed from $F_L$.  Since
$\ket{\varphi\psi}^L=2^{L/2} \calA_L (\ket\varphi\ket\psi)^L$,
\begin{equation*}
  F_L(\xi) 
  = 2^L \bigl( \bra{\varphi}\bra\psi \bigr)^L 
  \calA_L \hat F \calA_L
  \bigl( \ket\varphi\ket\psi \bigr)^L \;.
\end{equation*}
Due to the anti-symmetrisation between $\ket\varphi$ and $\ket\psi$
this function can be homogeneously extended to general
(non-orthonormal) $\ket\varphi$ and $\ket\psi$.  We choose
$\ket\varphi = \sum_{n=1}^N a_n \ket n$ and $\ket\psi = \sum_{n=1}^N
b_n \ket n$ with canonical basis vectors $\ket n$.  By differentiating
with respect to $a$, $b$ and their complex conjugates, all matrix
elements of $\calS_L\calA_L\hat F\calA_L\calS_L$ can be obtained.  
Using the symmetry (\ref{eq:F-sym}), we obtain $\calS_L\hat F$ and thus
also $\frac{2^L}{L+1}\calA_L\calS_L\hat F=\hat F$.

\section{Restrictions and direct products}
\label{app:representations}

We shall derive the relation between the restriction of representations
of $G=SU(N)$ to $H=\stab$ and the direct product of certain
representations used in section \ref{sec:harmonic}.  In this appendix,
we will allow for columns of height $N$, $J=[j_1,j_2,\ldots,j_N]$, in
diagrams describing representations of $SU(N)$.  These do not lead to
new representations, since representations differing only by $j_N$ are
unitarily equivalent, but this generalisation will make formulas much
simpler.  We embed $H$ into $SU(N)$ as
\begin{equation}
  \label{eq:embed}
  \begin{pmatrix}
    \ee^{\ii(N-k)\varphi} U' & 0 \\
    0 & \ee^{-\ii k \varphi} U''
  \end{pmatrix}
\end{equation}
where $U'\in SU(k)$ and $U''\in SU(N-k)$ and $\ee^{\ii\varphi}\in
U(1)$.  This shows that $H=[SU(k)\times SU(N-k)\times U(1)]/Z_{n}$
where $n$ %=k(N-k)/\gcd(k,N-k)$.  
is the least common multiple of $k$ and $N-k$.  Representations of $H$
can thus be considered as representations of $SU(k)\times SU(N-k)\times
U(1)$ that represent $Z_n$ trivially.  This fixes the charge $q$ of the
$U(1)$ factor $\ee^{\ii q\varphi}$ of the representation modulo $n$.
We will denote these representations as $(J',J'')_q$ where $J'$ and
$J''$ are symbols of $SU(k)$ and $SU(N-k)$ representations,
respectively, and $q$ is the charge of the $U(1)$ representation.

The restriction of an $SU(N)$ representation to $H$ can be written as
\begin{equation}
  \label{eq:restriction}
  J \bigr|_{H} = \bigoplus_{J',J''} m^J_{J',J''} \,
  (J',J'')_{(N-k)|J'|-k|J''|} \;.
\end{equation}
Here, $|J|=\sum_i j_i$ is the number of boxes in the diagram $J$ and we
assume that the diagrams have been chosen such that the total number
of boxes in $J'$ and $J''$ is the same as in $J$,
\begin{equation}
  \label{eq:box-sum}
  |J| = |J'| + |J''| \;.
\end{equation}
Note that the $U(1)$ representation is determined by the $SU(k)\times
SU(N-k)$ representation, so the multiplicities $m^J_{J',J''}$ are the
same as for the restriction from $SU(N)$ to the latter.  It is known
(\cite{Littlewood,Robinson}, also see \cite{Hagen-Macfarlane,Whippman})
that these can be obtained from the decomposition of the direct product
of the $SU(N)$ representations with diagrams $J'$ and $J''$,
\begin{equation}
  \label{eq:direct-prod}
  J' \otimes J''
  = \bigoplus_{J} m^J_{J',J''} \,
  J \qquad\text{in $SU(N)$.}
\end{equation}
Here, again, the restriction (\ref{eq:box-sum}) on the number of boxes
applies.  Note that in \eq~(\ref{eq:direct-prod}) $J'$ and $J''$ are
interpreted as $SU(N)$ representations while they are interpreted as
$SU(k)$ respectively $SU(N-k)$ representations in
\eq~(\ref{eq:restriction}).

We are interested in the case where the trivial representation of $H$
appears on the right-hand side of (\ref{eq:restriction}).  This means
that $J'$ and $J''$ only have columns of height $k$ and $N-k$,
respectively, $J'=[\mathbf{0}_{k-1},L',\mathbf{0}_{N-k-1}]$ and
$J''=[\mathbf{0}_{N-k-1},L'',\mathbf{0}_{k-1}]$ where
$\mathbf{0}_{k-1}$ stands for $k-1$ zero entries, etc.  In addition,
the $U(1)$ charge has to vanish, $(N-k)|J'|=k|J''|$.  Since $|J'|=L'k$
and $|J''|=L''(N-k)$, this implies $L'=L''\equiv L$, so that $J'$ and
$J''$ are complex conjugate representations of $SU(N)$.  We conclude
that a representation of $SU(N)$ contains the trivial representation
of $\stab$ if and only if it appears in the decomposition of the
direct product
\begin{equation}
  \label{eq:matrix-alg-app}
  M_L \equiv [\mathbf{0}_{k-1},L,\mathbf{0}_{N-k-1}]
  \otimes [\mathbf{0}_{N-k-1},L,\mathbf{0}_{k-1}]
\end{equation}
for some $L$.  The multiplicities are given by the multiplicities
$m^J_{[\mathbf{0}_{k-1},L,\mathbf{0}_{N-k-1}],
  [\mathbf{0}_{N-k-1},L,\mathbf{0}_{k-1}]}$ in the product.  Note that
the multiplicity does not depend on the number $j_N$ of columns of
height $N$ in the diagram $J$ chosen for a given representation.  This
means that a representation appears in $M_L$ if and only if $N
L\ge|J|$ where $J$ is the minimal diagram ($j_N=0$) of the
representation.  Therefore $M_{L}\subset M_{L'}$ if $L<L'$.

\subsubsection*{Decomposition of the direct product}
\label{app:decomposition}

For illustrational purposes, we will explicitly perform the
decomposition of the direct product (\ref{eq:matrix-alg-app}) into
irreducible representations.  We can assume $k\le N-k$ since
$G^N_k\cong G^N_{N-k}$.  The decomposition is achieved by Young
diagram techniques.  Recall the rules for decomposing the direct
product of two irreducible representations of $SU(N)$
\cite{Sternberg}:
\begin{enumerate}
\item Label each box in the second diagram by its row number.
\item Attach all boxes with 1's to the first diagram, then all
  boxes with 2's and so on, such that
  \begin{enumerate}
  \item at all stages the intermediate diagram corresponds to an
    irreducible representation of $SU(N)$, i.e.{\ }all columns start in
    the first row and are connected and the height of the columns
    monotonically decreases from left to right,
  \item no column contains any number more than once, and
    \label{once}
  \item\label{order} when counted from the right, the $n$-th $i$ does
    not appear before the $n$-th $i-1$.
  \end{enumerate}
\end{enumerate}
Applying these rules to our case, we have to attach boxes with $L$
copies of each of the numbers $1,\dots,k$ to a rectangle of height
$N-k$ and width $L$.  This is indicated in figure~\ref{fig:product}.
{\let\x=\times\def\o{\cdot\cdot}
\begin{figure}[tbp]
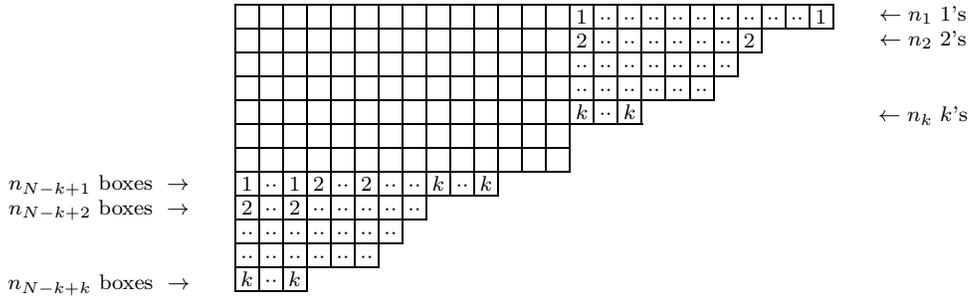

  \centering
  \begin{equation*}
    \vcenter{\hbox{\scriptsize$\begin{matrix}
        \\[6.6em]
        n_{N-k+1} \text{ boxes }\rightarrow \\
        n_{N-k+2} \text{ boxes }\rightarrow \\
        \\
        \\
        n_{N-k+k} \text{ boxes }\rightarrow
      \end{matrix}
      \qquad
      \Yautoscale1\young(\ \ \ \ \ \ \ \ \ \ \ \ \ \ 1\o\o\o\o\o\o\o\o\o1,\ \ \ \ \ \ \ \ \ \ \ \ \ \ 2\o\o\o\o\o\o2,\ \ \ \ \ \ \ \ \ \ \ \ \ \ \o\o\o\o\o\o\o,\ \ \ \ \ \ \ \ \ \ \ \ \ \ \o\o\o\o\o\o,\ \ \ \ \ \ \ \ \ \ \ \ \ \ k\o k,\ \ \ \ \ \ \ \ \ \ \ \ \ \ ,\ \ \ \ \ \ \ \ \ \ \ \ \ \ ,1\o12\o2\o\o k\o k,2\o2\o\o\o\o\o,\o\o\o\o\o\o\o,\o\o\o\o\o\o,k\o k)
      \qquad \begin{matrix}
        \leftarrow n_1 \text{ 1's}\\
        \leftarrow n_2 \text{ 2's} \\
        \\
        \\
        \leftarrow n_k \text{ $k$'s} \\[7.3em]
      \end{matrix}$}}
  \end{equation*}
  \caption{Decomposition of the tensor product $[\mathbf{0}_{N-k-1},L,\mathbf{0}_{k-1}]\otimes [\mathbf{0}_{k-1},L,\mathbf{0}_{N-k-1}]$.}
  \label{fig:product}
\end{figure}}
We have already anticipated some facts about the resulting distribution
of boxes, which we will explain now.  The number of 1's in the first row
has been denoted by $n_1$.  The remaining $L-n_1$ 1's have to be in the
$N-k+1$-st row.  Denoting the number of 2's in the second row by $n_2$,
there can be at most $n_1-n_2$ 2's in row $N-k+1$, because of rule
\ref{order}.  There must thus be at least $L-n_1$ 2's in row $N-k+2$.
However, owing to rule \ref{once}, the number of 2's in that row is at
most $L-n_1$.  Therefore, there have to be exactly $n_1-n_2$ 2's in row
$N-k+1$ and $L-n_1$ 2's in row $N-k+2$.  By the same reasoning, we find
that the number of $i$'s in row $N-k+l$ for $l>0$ is equal to the
number of $i-l+1$'s in the row $N-k+1$ which is in turn equal to
$n_{i-l}-n_{i-l+1}$ if $i\ge l$ and 0 if $i<l$ (we have put
$n_0\equiv L$).  Adding up, we find for the number of boxes in row
$N-k+l$,
\begin{equation}
  n_{N-k+l} = \sum_{i=l}^k (n_{i-l} - n_{i-l+1} ) = L- n_{k-l+1}
  \quad\text{for}\quad l=1,\ldots,k \;.
\end{equation}
Graphically, this means that the subdiagram of columns $L+1,
L+2,\ldots$ combines, after a rotation by $\pi$, with the remainder of
the diagram to a rectangle of width $L$ and height $N$.  In terms of
symbols $J=[j_1,\ldots,j_N]$, where $j_i$ is the number of columns of
height $i$, this implies
\begin{equation}
  \label{eq:harmonic-app}
  J = \begin{cases}
    [m_1,\ldots,m_k,0,\ldots,0,m_k,\ldots,m_1] & \text{if } 2k<N
    \;, \\{}
    [m_1,\ldots,m_{k-1},2m_k,m_{k-1},\ldots,m_1] & \text{if } 2k=N 
  \end{cases}
\end{equation}
with $m_i=n_i-n_{i+1}$ where $n_{k+1}\equiv0$.  All non-negative
values of $m_i$ satisfying $\sum_{i=1}^k m_i=n_1\le L$ occur.
Furthermore, each diagram can be obtained in only one way, since it is
determined by the numbers $n_i$ ($i=1,\dots,k$).  So all
multiplicities equal 1.

\section{Symmetric group}
\label{app:sym}

In this appendix, we shall provide a proof of the factorisation
property (\ref{eq:K-factor}).  On the way, we will recall some facts
about representations of the symmetric group and the associated
projectors $\Pchar_J$ used in the text.  These projectors can be
considered as elements of the group algebra $\alg\equiv\spanop
S_n=\{A=\sum_{\pi\in S_n} A_\pi \pi \vert A_\pi\in\R\}$, the set of
formal linear combinations of group elements.  The vector space $\alg$
carries a representation of the algebra $\alg$ whose action is given
by left multiplication.  This representation restricts to a
representation of the subgroup $S_n$ of $\alg$.  It is usually called
the regular representation.  Sometimes it is convenient to identify
$\alg$ with the algebra $F(S_n)$ of functions on the group $S_n$ by
setting $A(\pi)\equiv A_\pi$.  The action of a group element $\pi$ is
then given by $(\pi A)(\sigma)=A(\pi^{-1}\sigma)$.  $F(S_n)$ can in
fact be considered as the dual vector space of $\alg$ since each
function on the group can be linearly and uniquely extended to a
function on $\alg$.  The identification of $\alg$ with its dual space
can be obtained from the inner product
\begin{equation}
  \label{eq:inner-prod}
  \langle A,B \rangle \equiv \sum_{\pi\in S_n} A_\pi B_\pi 
  = \frac{1}{n!} \tr (A^\T B) 
\end{equation}
by putting $A(B)=\langle A,B\rangle$. The trace in
\eq~(\ref{eq:inner-prod}) is over the regular representation and we
have set $A^\T=\sum_\pi A_\pi \pi^{-1}$.

Of particular importance are the central elements of $\alg$, that are
invariant under conjugation with any group element $\pi\in S_n$, $A=\pi
A\pi^{-1}$.  They correspond to class functions, i.e.{\ }functions that
depend only on the conjugacy class of their argument.  An orthogonal
basis in the subspace of class functions is given by the characters
$\chi_J$ associated with the irreducible representations $J$
of $S_n$,
\begin{equation}
  \label{eq:character-orth}
  \langle \chi_J, \chi_{J'} \rangle 
  = n! \, \delta_{J J'} \;.
\end{equation}
So every class function can be expanded as
\begin{equation}
  \label{eq:character-expansion}
  A = \smash{\sum_J} A_J \chi_J
  \qquad\text{with}\qquad
  A_J = \frac{1}{n!} \langle\chi_J,A\rangle\;.
\end{equation}
To each $A\in\alg$, one can associate a central element $\overline A$
by averaging with respect to conjugation,
\begin{equation}
  \label{eq:class-average}
  \overline A \equiv \frac{1}{n!} \sum_{\pi\in S_n} \pi A \pi^{-1}
  = \frac{1}{n!} \sum_J \langle\chi_J,A\rangle \, \chi_J 
\end{equation}
where we have used (\ref{eq:character-expansion}) and the invariance
of $\chi_J$ under conjugation.

The regular representation is in general reducible.  It contains each
irreducible representation $J$ with a multiplicity that is given
by the dimension $d_J$ of the representation.  The component
containing all copies of an irreducible representation $J$ can be
obtained as the image of the symmetric projection operator introduced
in \eq~(\ref{eq:decomp-def}), in the dual picture,
\begin{equation}
  \label{eq:char-proj}
  \Pchar_J = \frac{d_J}{n!} \chi_J \;.
\end{equation}
The decomposition of unity
\begin{equation}
  \label{eq:char-decomp}
  1 = \sum_{J} \Pchar_J
\end{equation}
provides a decomposition of $\alg$ into orthogonal subspaces \cite{BR}.

Now we will show how the averaged tensor product of two projectors
$\Pchar_{J_1}$ and $\Pchar_{J_2}$ onto irreducible representations of
$S_{n_1}$ and $S_{n_2}$ can be expressed in terms of irreducible
projectors.  $\Pchar_{J_1}\otimes \Pchar_{J_2}$ can be extended to
$\alg$ where $n=n_1+n_2$.  By (\ref{eq:class-average}), we have
\begin{equation}
  \label{eq:YY-expansion}
  \overline{\Pchar_{J_1} \otimes \Pchar_{J_2}} 
  = \sum_J a_J \chi_J
\end{equation}
with
\begin{equation}
  \label{eq:a-J}
  a_J = \frac{1}{n!} \langle
  \chi_J , \Pchar_{J_1}\otimes \Pchar_{J_2} \rangle \;.
\end{equation}
The restriction of $\chi_J$ to $(\pi,\sigma)\in S_{n_1}\times
S_{n_2}$ decomposes into irreducible characters as
\begin{equation}
  \chi_J(\pi,\sigma)
  = \sum_{J_1,J_2} c^J_{J_1 J_2}
  \chi_{J_1}(\pi) \chi_{J_2}(\sigma)
\end{equation}
where $c^J_{J_1 J_2}\in\Z$ are multiplicities or
Clebsch-Gordan coefficients.  With (\ref{eq:character-orth}) we get
\begin{equation}
  a_J 
  = \frac{1}{n!} \sum_{J_1',J_2'} 
  c^J_{J_1' J_2'} \, 
  \langle \chi_{J_1'} , \Pchar_{J_1} \rangle \, 
  \langle \chi_{J_2'} , \Pchar_{J_2} \rangle
  = \frac{d_{J_1} d_{J_2}}{n!}  c^J_{J_1 J_2}
\end{equation}
and therefore
\begin{equation}
  \label{eq:YY-class}
  \overline{ \Pchar_{J_1} \otimes \Pchar_{J_2}} 
  = \sum_J \frac{d_{J_1} d_{J_2}}{n!}
    c^J_{J_1 J_2} \chi_{J}
  = \sum_J \frac{d_{J_1} d_{J_2}}{d_J}
    c^J_{J_1 J_2} \overline{\Pchar_J} \;.
\end{equation}
By iteration, this result can be generalised to multiple products,
\begin{equation}
  \label{eq:Y-mult-class}
  \frac{1}{d_{J_1} \cdots d_{J_m}}
  \overline{ \Pchar_{J_1} \otimes \cdots \otimes \Pchar_{J_m}}
  = \sum_J c^J_{J_1 \dots J_m} \frac{1}{d_J} 
  \overline{\Pchar_J} \;.
\end{equation}
Note that symmetrisation with respect to $S_n$ implies symmetrisation
with respect to $S_{n'}\subset S_n$,
\begin{equation}
  \label{eq:sym-iter}
  \overline{A \otimes B \otimes C}
  = \overline{A \otimes \overline{B \otimes C}} \;.
\end{equation}
Now we can prove \eq~(\ref{eq:K-factor}).  The right-hand side of this
equation can be written as a single trace like in
\eq~(\ref{eq:K-J-explicit}) but with $\Pchar_J$ replaced by
$\Pchar_{[l]}\otimes \Pchar_{[0,1]}^{\otimes m}$.  Owing to the
symmetric tensors $S$ and $T$ all factors in the trace except
$\Pchar_{[l]}\otimes \Pchar_{[0,1]}^{\otimes m}$ are symmetric under
conjugation, so $\Pchar_{[l]}\otimes \Pchar_{[0,1]}^{\otimes m}$ can be
replaced by its symmetrised version $\overline{\Pchar_{[l]}\otimes
  \Pchar_{[0,1]}^{\otimes m}}$.  Now we can insert
\eq~(\ref{eq:Y-mult-class}).  The only term on the right-hand side that
does not vanish when projected by $\calP^{\otimes n}$ to a
representation of $SU(2)$ is $J=[l,m,0,\dots]$ with multiplicity $1$.
The dimensions of the representations $[l]$ and $[0,1]$ (of the
symmetric group) are 1, while the dimension of $[l,m]$ appearing in the
denominator of (\ref{eq:Y-mult-class}) just cancels that on the
right-hand side of (\ref{eq:K-factor}), so we obtain the left-hand
side.

\section{Projection of multiple derivatives}
\label{app:multi-der}

We compute the product of multiple (anti-)holomorphic derivatives of
$\omega_L$ and $\omega_L'$ projected to the representation
$([l,m]^*,[l,m])$ of the stability group,
\begin{equation}
  X_{l,m}^{(L)} 
  \equiv (\partial_{A_1} \cdots \partial_{A_{l+2m}} \omega^L ) \, 
  K_{[l,m]}^{A_1\dots A_{l+2m},B_1\dots B_{l+2m}} \,
  (\partial_{B_1} \cdots \partial_{B_{l+2m}} \omega^{\prime L} ) \;.
\end{equation}
By \eq~(\ref{eq:multi-hol}) and since $K_{[l,m]}$ contains the
projector $K$, the derivatives in this equation are really covariant
derivatives, holomorphic ones acting on $\omega'$ and anti-holomorphic
ones on $\omega$. Equation~(\ref{eq:K-factor}) implies that
$K_{[l,m]}$ can be replaced by $d_{[l,m]} K_{[l]}\otimes
K_{[0,1]}^{\otimes m}$,
\begin{equation}
  X_{l,m}^{(L)} 
  = d_{[l,m]} \, ( \partial^{l+2m} \omega^L ) \, \bigl( K_{[l]}
  \otimes K_{[0,1]}^{\otimes m} \bigr) \,
  ( \partial^{l+2m} \omega^{\prime L} ) 
\end{equation}
where we have introduced an index-free notation.  Using the second
equality of \eq~(\ref{eq:d2omega=domega2}), we find
\begin{equation}
  K_{[0,1]}^{A B,C D} \partial_C \partial_D \omega^{L}
  = \frac{L(L+1)}{2} \omega^{L-1}
  K_{[0,1]}^{A B,C D} \partial_C \partial_D \omega
\end{equation}
which iterates to
\begin{equation}
  ( K_{[0,1]} \partial\partial )^m \omega^L 
  = \frac{L! (L+1)!}{(L-m)! (L+1-m)!} \,
  \omega^{L-m}
  \left( \tfrac12 K_{[0,1]} \partial\partial\omega \right)^{m}
\end{equation}
because the triple derivative of $\omega$ vanishes.  The first
equality of \eq~(\ref{eq:d2omega=domega2}) implies that
$K_{[l]}\partial^l\omega^n$ contains only single derivatives of
$\omega$, whence
\begin{multline}
  \bigl( K_{[l]} \otimes K_{[0,1]}^{\otimes m} \bigr) 
  \partial^{l+2m} \omega^L
  = \bigl( K_{[l]} \otimes K_{[0,1]}^{\otimes m} \bigr) 
  \bigl[ \partial^{l} (K_{[0,1]} \partial\partial)^m \omega^{ L} \bigr] \\
  = \frac{L! (L+1)!}{(L-l-m)! (L+1-m)!} \,
  \omega^{ L-l-m}
  \bigl[ K_{[l]} (\partial\omega)^{l} \bigr]
  \left[ \tfrac12 K_{[0,1]} (\partial\partial\omega) \right]^{m}
\end{multline}
where, in the first step, we have used \eq~(\ref{eq:KKdK}).  Since a
similar equality holds for anti-holomorphic derivatives, and $K_{[l]}$
and $K_{[0,1]}$ are projectors, we find
\begin{equation}
  \label{eq:X-res}
  \begin{split}
    X_{l,m}^{(L)}
    &= d_{[l,m]} \, \left( \frac{L! (L+1)!}{(L-l-m)!
        (L+1-m)!} \right)^2
    (\omega\omega')^{ L-l-m} \\
    &\quad \times
    \bigl[ (\partial\omega)^l K_{[l]} (\partial\omega')^{l} \bigr]
    \left[ \tfrac14 (\partial\partial\omega) K_{[0,1]}
      (\partial\partial\omega') \right]^{m} \;.
  \end{split}
\end{equation}

\end{document}